\documentclass[twoside,a4paper,10pt]{proceedings}
\usepackage{graphicx}
\usepackage{natbib}

\def\1{\c{c}}
\def\2{\c{C}}
\def\3{\.{I}}
\def\4{\"{a}}
\def\5{{\i}}
\def\6{$\beta$}
\def\7{\"{o}}
\def\8{\"{O}}
\def\9{\c{s}}
\def\0{\c{S}}
\def\*{\"{u}}
\def\;{\u{g}}
\def\:{\u{G}}

\begin{document}

\pagenumbering{arabic}
\pagestyle{myheadings}
\thispagestyle{empty}
\vspace*{-1cm}
{\flushleft\includegraphics[width=3cm,viewport=0 -30 200 -20]{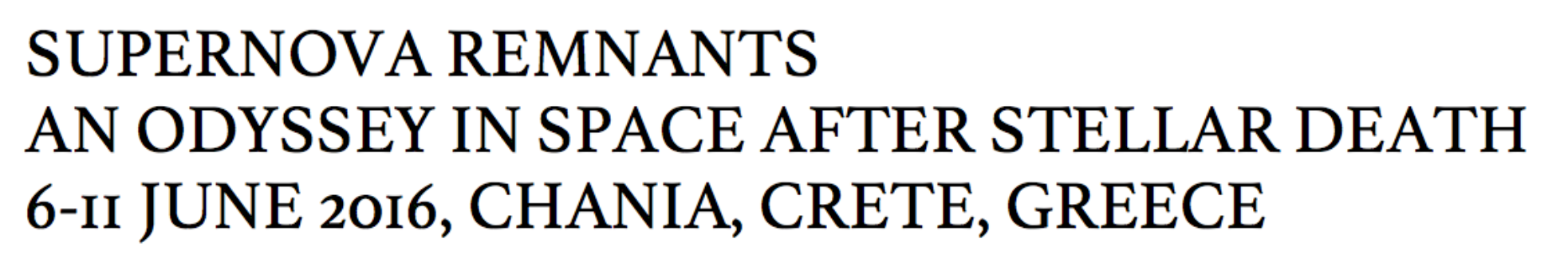}}
\vspace*{0.2cm}

\begin{flushleft}
{\bf {\LARGE
%%% TITLE of the paper. 
Searching for the Time Variation in Supernova
 Remnant RX J1713.7$-$3946
}\\
\vspace{0.3cm}
%%% Include here the LIST OF AUTHORS.
%%% Note that the last author has to be preceeded by an AND.
Aytap Sezer$^1$,
Ryo Yamazaki$^2$,
Xiaohong Cui$^3$,
Aya Bamba$^4$
and Yutaka Ohira$^2$
}\\
\vspace*{0.2cm}
%
%%% AFFILIATIONS LIST.
%%%Note that one affiliation per line.
{\footnotesize
$^{1}$Harvard-Smithsonian Center for Astrophysics, 60
Garden Street, Cambridge, MA 02138, USA \\
$^{2}$Department of Physics and Mathematics, Aoyama Gakuin University, 5-10-1, Fuchinobe, Sagamihara 252-5258, Japan \\
$^{3}$National Astronomical Observatories, Chinese Academy of Sciences
20A Datun Road, Chaoyang District, Beijing, 100012, China\\
$^{4}$Department of Physics, The University of Tokyo, 7-3-1 Hongo,
Bunkyo-ku, Tokyo 113-0033, Japan\\
}
\end{flushleft}

%%% HEADINGS
\markboth{
Searching for the Time Variation in Supernova Remnant RX J1713.7$-$3946
}
{
Sezer et al.
}
\thispagestyle{empty}

\vspace{-1.0cm}
%%% ABSTRACT
\section{Abstract}
\vspace{-0.3cm}
Supernova Remnant RX J1713.7$-$3946 emits synchrotron X-rays and very high energy
 $\gamma$-rays. Recently, thermal X-ray line emission is detected from ejecta plasma.
 CO and H\,{\sc i} observations indicate that a highly inhomogeneous medium surrounding the SNR.
 It is interacting with dense molecular clouds in the northwest and the southwest of the
 remnant. The origin of the $\gamma$-ray emission from RX J1713.7$-$3946 is still uncertain.
 Detection of rapid variability in X-ray emission from RX J1713.7$-$3946 indicates the magnetic
 field $B$ $\sim$ mG. In this work, we investigate the time variation in X-ray flux, luminosity
 and photon index of RX J1713.7$-$3946. For this investigation, we study the northwest part of
 the remnant using {\it Suzaku} data in 2006 and 2010. We present preliminary results based on
 our analysis and interpretations about these X-ray time variability.

%%%INTRODUCTION
\vspace{-0.5cm}
\section{Introduction}
\vspace{-0.3cm}
X-ray emission of supernova remnant
(SNR) RX J1713.7$-$3946 (G347.3$-$0.5) is
dominated by synchrotron radiation (e.g., \citealt{Ko97, Sl99}). Recently, \citet {Ka15} detected thermal X-ray emission from
this SNR using {\it XMM-Newton} and {\it Suzaku} data. \citet {Sa10} showed that the synchrotron X-ray emission is enhanced around the CO clumps in the northwest (NW) region of this remnant. 

The time variation of synchrotron X-rays was discovered in compact regions of the NW shell of RX J1713.7$-$3946 \citep {Uc07}. They proposed that the magnetic field in these regions is amplified to 1 mG. The X-ray variability provides information about average magnetic field condition. In this work, we investigate the time variation in X-ray flux
and photon index of the NW part of RX J1713.7$-$3946 using {\it Suzaku} data in 2006 and 2010.
We describe these observations and data analysis in Section 3. We discuss the preliminary results of our data analysis in Section 4.

%%% X-RAY ANALYSIS
\vspace{-0.5cm}
\section{Observation and Analysis}
\vspace{-0.2cm}
The NW region of RX J1713.7$-$3946
observed with X-ray Imaging Spectrometer \citep[XIS;][]{Ko07}
on 2006 September (Obs ID:
501063010) and 2010 February (Obs
ID:504027010) for $\sim$18.4 ks and
61.5 ks, respectively. For our analysis we used the {\sc HEAsoft} package
(version 6.16) and {\sc xspec} version 12.9.0
 \citep {Ar96}. For the spectral analysis we generated the response matrix and auxiliary files (RMFs and ARFs) using
{\sc xisrmfgen} and {\sc
xissimarfgen} \citep {Is07}, respectively. 

Figure 1 shows the XIS images of 2006 and 2010 observations.
We extracted X-ray spectra from a circular region with a radius 2.6 arcmin for both 2006 and 2010 data indicated in Figure 1.
We subtracted the non-X-ray background (NXB) from both observations. The NXB spectra were made by using {\sc xisnxbgen} \citep {Ta08}.
We fitted both spectra with a power-law (PL) model modified by an absorption model \citep[TBABS:][]{Wi00}. 
For this fitting, the column density ($N_{\rm H}$), the photon index ($\Gamma$) and the normalization of PL component were left as free parameters. The results of two observations are given in Table 1, and the XIS spectra are
shown in Figure 2.

\begin{figure*}
\centering \vspace*{1pt}
\includegraphics[width=0.45\textwidth]{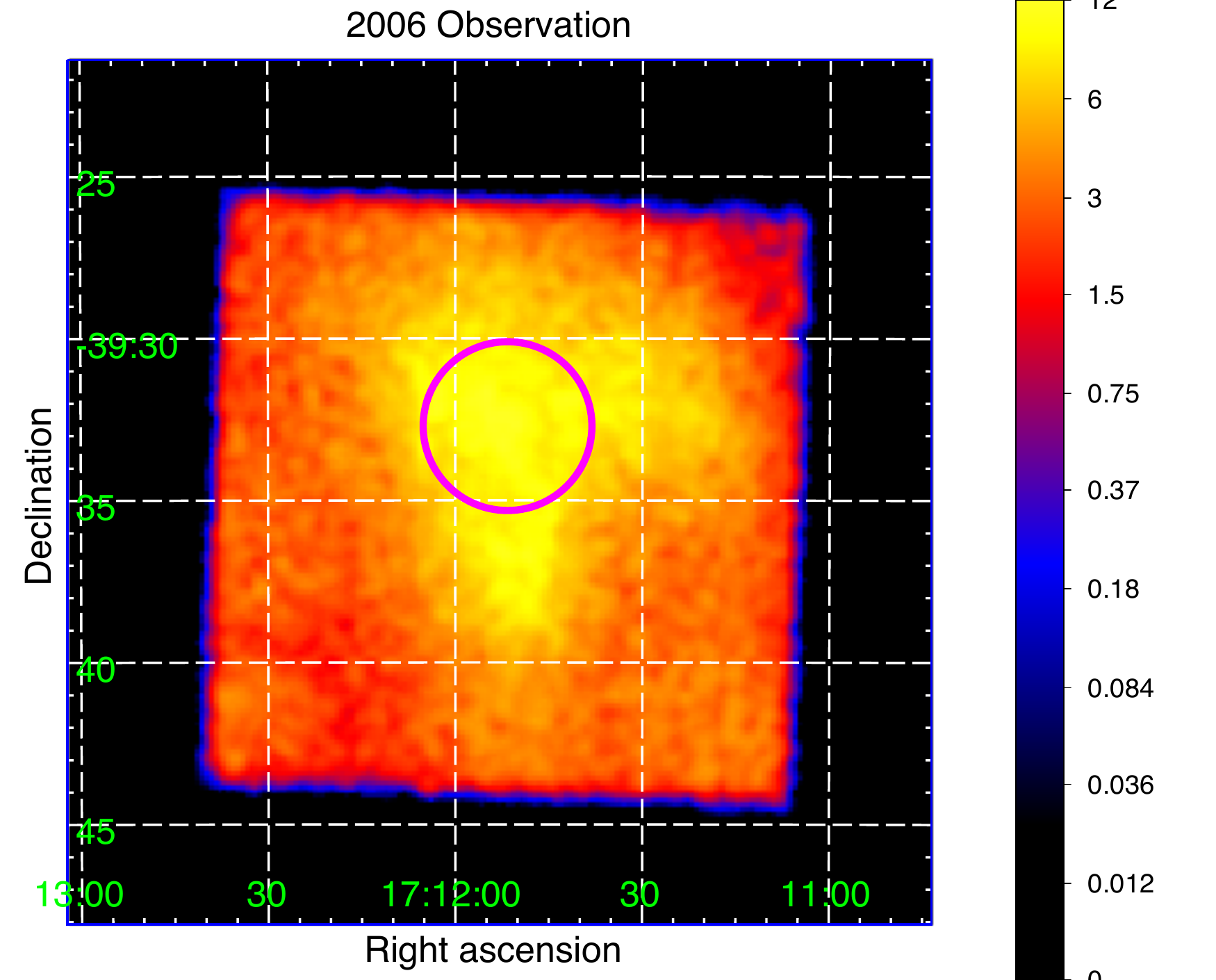}
\includegraphics[width=0.46\textwidth]{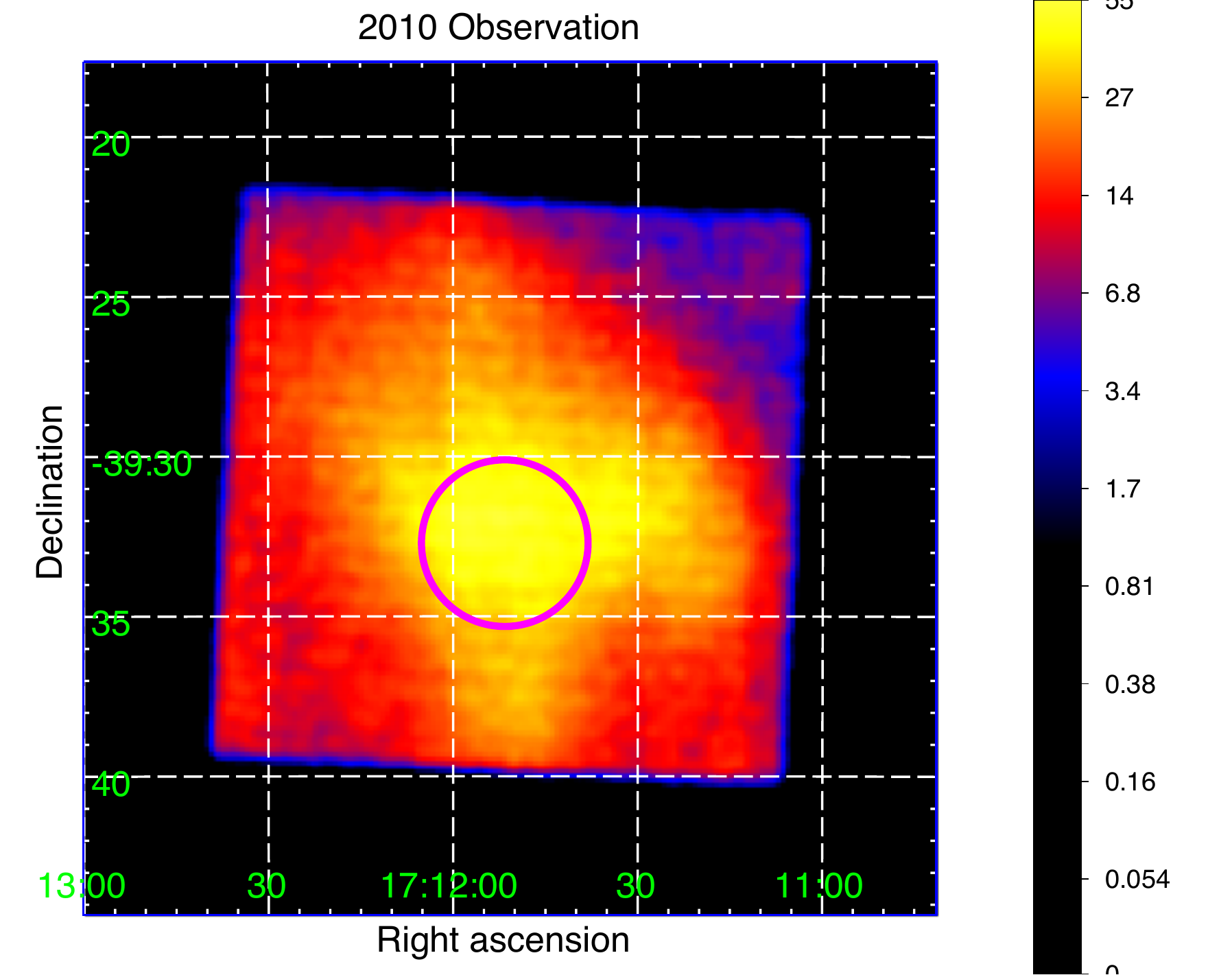}
\caption{{\it Suzaku} XIS image of RX J1713.7$-$3946 in the 0.3$-$10.0 keV
 energy band, 2006 observation (left) and 2010 observation (right). The spectral analysis
regions are shown by circles.}
\end{figure*}

\begin{figure*}
\centering
  \vspace*{1pt}
\includegraphics[width=0.48\textwidth]{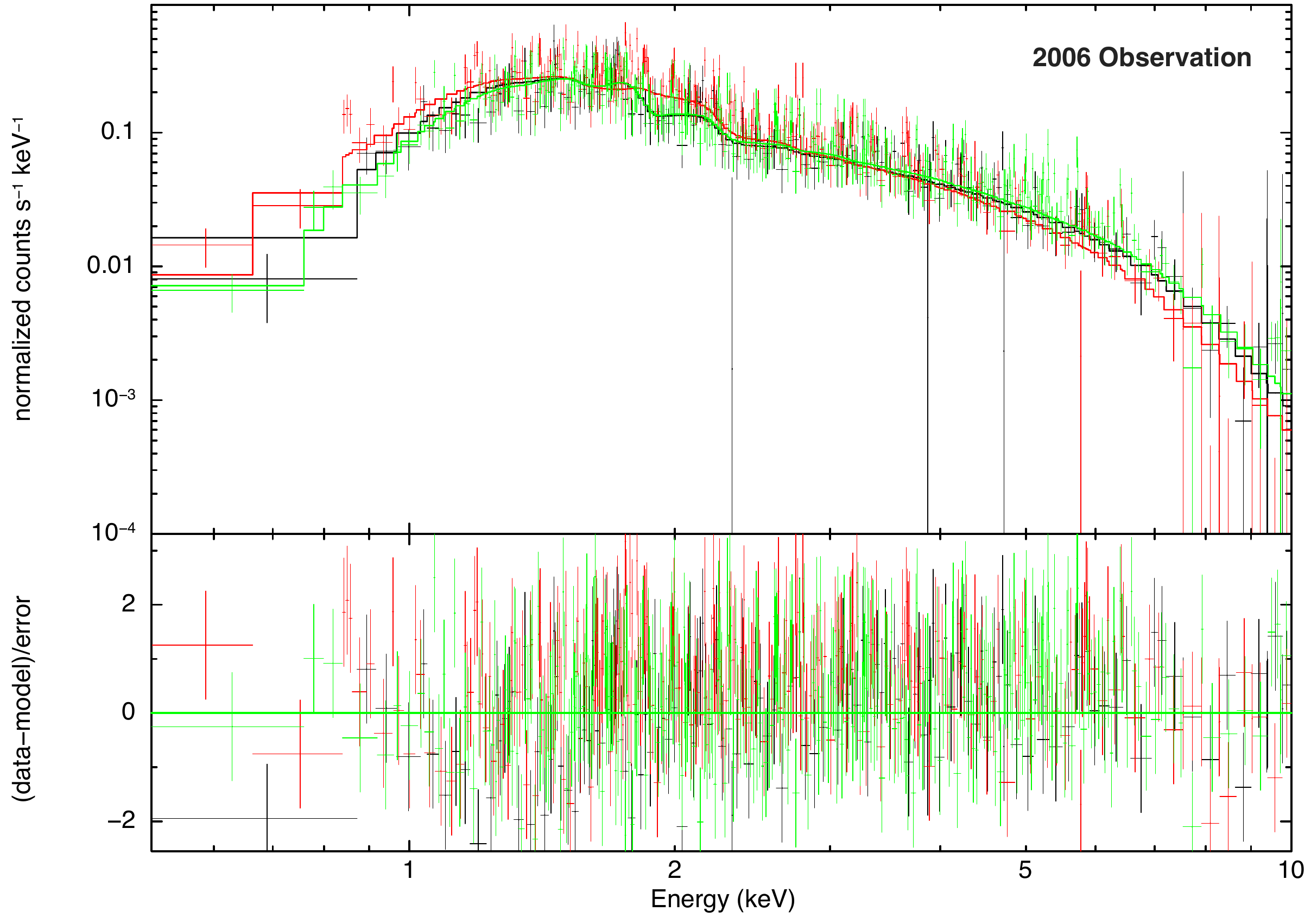}
\includegraphics[width=0.45\textwidth]{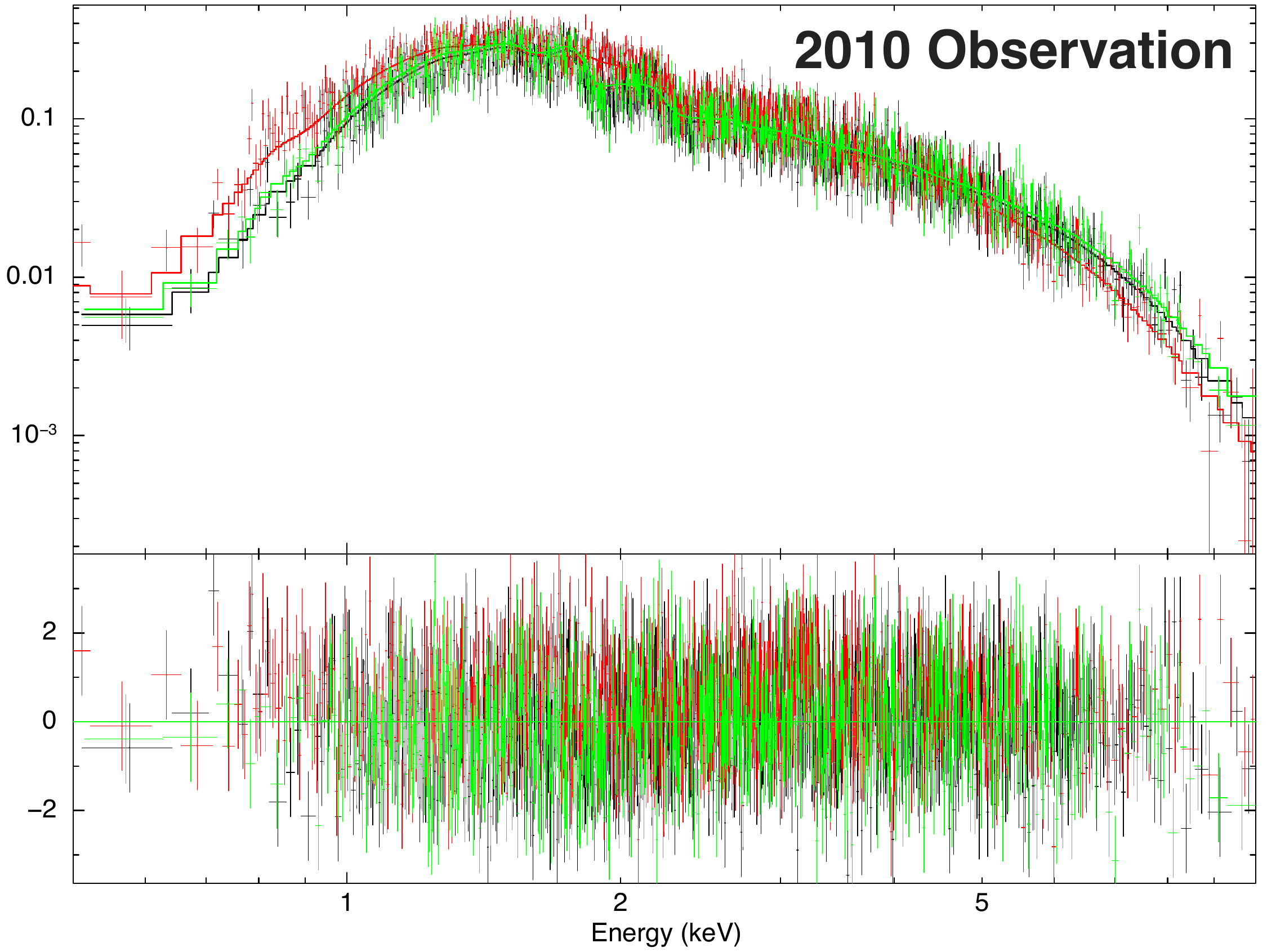}
  \caption{ {\it Suzaku} XIS spectra of 2006 observation (left) and 2010 observation (right) of RX J1713.7$-$3946 in
 the 0.5$-$10.0 keV energy band fitted with an absorbed PL model.}
\end{figure*}

\begin{table*}
\caption{Spectral parameters of the NW region of RX J1713.7$-$3946. All uncertainties correspond to the 90\% confidence level.}
 \begin{minipage}{160mm}
 \begin{center}
\begin{tabular}{@{}lllll@{}}
    \hline
 \hline
    \\
&Component& Parameter  &  2006 Observation & 2010 Observation \\
  \hline
  \\
&TBABS& $N_{\rm H}$ ($\times10^{22}$ cm$^{-2}$)  &  $0.83^{+0.05}_{-0.05}$ & $0.88^{+0.02}_{-0.01}$\\
&PL & Photon Index  &  $2.43^{+0.03}_{-0.04}$  & $2.39^{+0.02}_{-0.02}$ \\
&& Flux ($\times10^{-11}$ erg cm$^{-2}$ s$^{-1}$)  &  $1.19^{+0.01}_{-0.01}$ &$1.41^{+0.01}_{-0.01}$\\
\\[-0.7ex]
\hline
&& $\chi^{2}$/dof &  5496.7/7801 & 8406/7668 \\
  \hline
\end{tabular}
\end{center}
\end{minipage}
\end{table*}

%%%Discussion
\vspace{-0.6cm}
\section{Discussion}
\vspace{-0.2cm}

As seen in Table 1, our preliminary results show that the flux increases in 4 years. 
There is no variation of the photon index within error range. There are 
a few reasons for the flux increasing. One is that the relativistic electrons emitting synchrotron X-rays are freshly accelerated or reaccelerated. Other possibilities are increasing the emission volume or the amplification of the magnetic field. In any of these cases, the synchrotron X-ray emission becomes brighter.

The angular size of our analyzed region is about $0.03^\circ$, which is 
comparable to the angular resolution of HESS and CTA. Therefore, a variability in TeV gamma-ray could be observed. 

It is expected that the diffusive shock acceleration (DSA) occurs at
the shock front of SNR. Our observation region is the downstream
region in the RX J1713.7$-$3946. Our initial results indicate that other
kinds of acceleration mechanism than DSA, suggesting, e.g., the 2nd
order Fermi acceleration (in other words, the turbulent acceleration).

Usually, it is generally expected that the flux is decreasing as SNR
ages. However, our result shows that the flux is increasing, which
implies that the turbulence becomes stronger. Then, the turbulent
acceleration of electrons occurs. Our question is how the turbulence
becomes stronger. A possible solution is that the shock-cloud
interaction. This SNR is surrounded by many molecular clouds (MCs) \citep {Sa13}. Our
spectral region (NW) is near a large MC, Clump D, which is shown in
Figure 5 of \citet {Sa13}. Clump D excited strong turbulence,
which accelerate electrons in the downstream region.

In the next step of our search, we will investigate the larger spectral region in the same field of view to test our result.

% Do not delete the next line
\small
%%% ACKNOWLEDGMENTS
% Comment the following line if you do not have acknowledgments.
\vspace{-0.5cm}
\section*{Acknowledgments}   
\vspace{-0.3cm}
AS is supported by the Scientific and Technological Research Council of Turkey (T\"{U}B\.{I}TAK) through the B\.{I}DEB-2219 fellowship program. RY is supported in part by grant-in-aid from the Ministry of Education, Culture, Sports, Science, and Technology (MEXT) of Japan, No. 15K05088.

%%% BIBLIOGRAPHY

%%% END OF DOCUMENT
\end{document}